\documentclass[11pt,a4paper]{article}

\usepackage{jcappub}
\usepackage{epsf}  
\usepackage{eucal} 
\usepackage{amssymb}
\usepackage{amsmath} 
\usepackage{graphicx} 
\usepackage{bm}
\usepackage{dcolumn} 
\usepackage{epsfig} 
\usepackage{multirow}
\usepackage{hyperref}
\usepackage{color}
\usepackage{axodraw4j}

\begin{document}

\title{Delaying the waterfall transition in warm hybrid inflation}

\author[a]{Mar Bastero-Gil,}
\emailAdd{mbg@ugr.es}
\affiliation[a]{Departamento de F\'{\i}sica Te\'orica y del Cosmos, Universidad de Granada, Granada-18071, Spain}

\author[b]{Arjun Berera,}
\emailAdd{ab@ph.ed.ac.uk} 
\affiliation[b]{SUPA, School of Physics and Astronomy, University of Edinburgh, Edinburgh, EH9 3JZ, United Kingdom}

\author[b]{Thomas P. Metcalf}
\emailAdd{t.p.metcalf@ed.ac.uk}

\author[c]{and \\Jo\~ao G. Rosa}
\emailAdd{joao.rosa@ua.pt} 
\affiliation[c]{Departamento de F\'{\i}sica da Universidade de Aveiro and I3N, Campus de Santiago, 3810-183 Aveiro, Portugal}

\date{\today}


\abstract{
We analyze the dynamics and observational predictions of supersymmetric hybrid inflation in the warm regime, where dissipative effects are mediated by the waterfall fields and their subsequent decay into light degrees of freedom. This produces a quasi-thermal radiation bath with a slowly-varying temperature during inflation and further damps the inflaton's motion, thus prolonging inflation. As in the standard supercooled scenario, inflation ends when the waterfall fields become tachyonic and can no longer sustain a nearly constant vacuum energy, but the interaction with the radiation bath makes the waterfall fields effectively heavier and delays the phase transition to the supersymmetric minimum. In this work, we analyze for the first time the effects of finite temperature corrections and SUSY mass splittings on the quantum effective potential and the resulting dissipation coefficient. We show, in particular, that dissipation can significantly delay the onset of the tachyonic instability to yield 50-60 e-folds of inflation and an observationally consistent primordial spectrum, which is not possible in the standard supercooled regime when inflation is driven by radiative corrections.
}



\maketitle



\section{Introduction}

One of the most important problems in modern cosmology is the embedding of inflationary physics within a consistent quantum field theory framework that describes particle interactions at high energies. This is particularly significant given the requirement of a `graceful exit' from inflation into the standard Hot Big Bang evolution, after a sufficiently long period of accelerated expansion smooths out any primordial inhomogeneities and curvature, leaving behind a frozen spectrum of small density perturbations that later seed the temperature anisotropies in the Cosmic Microwave Background and the observed Large Scale Structure.

Hybrid inflation models are amongst the most promising avenues of research in this quest for a well defined particle physics description of inflation. Originally proposed by Linde \cite{Linde:1993cn}, these models combine the attractive features of inflationary scenarios driven by a slowly rolling scalar field and the original GUT constructions with spontaneous symmetry breaking and associated phase transitions. In particular, in hybrid models the scalar inflaton field is coupled to one or more additional scalar fields with a Higgs-like potential, known as the waterfall field(s).  For large inflaton values, these fields become heavy and are kept at a metastable minimum with a non-zero vaccum energy that can sustain accelerated expansion. As the inflaton rolls slowly down its potential, it will eventually reach a critical value below which this minimum becomes unstable, with both the inflaton and the waterfall field(s) being quickly driven to the true ground state. This occurs via a second order phase transition, thus overcoming the problem of reheating the universe through bubble nucleation and collisions. 

While the inflaton may be taken as a singlet field, preventing potentially large radiative corrections that destroy the required flatness of the inflaton potential, the waterfall fields may be charged under the Standard Model or GUT gauge groups, thus interacting with ordinary matter and gauge particles and reheating the universe during the phase transition. Hybrid inflation models of this form are, moreover, ubiquitous in supersymmetric (SUSY) extensions of the Standard Model, with both F- and D-term SUSY breaking models in global SUSY and supergravity having been discussed in the literature \cite{Dvali:1994ms, Copeland:1994vg, Linde:1997sj, Senoguz:2003zw, Senoguz:2004vu, BasteroGil:2006cm, Garbrecht:2006az, urRehman:2006hu, Pallis:2009pq, Rehman:2009nq, Rehman:2009yj, Civiletti:2013cra, Pallis:2013dxa}. Interesting implementations may also be found within the context of string/M-theory, in particular multiple D-brane constructions, where an inter-brane distance modulus plays the role of the inflaton and the waterfall field(s) are associated with the ground states of strings stretched between different brane and anti-brane stacks \cite{Burgess:2001fx}. These states become tachyonic below a critical inter-brane distance of the order of the fundamental string length, triggering a waterfall transition that ends with the annihilation of opposite charge D-branes.

The coupling between the inflaton and the waterfall fields plays a crucial role in hybrid inflation models, since it not only determines the mass of the waterfall fields and hence the stability of the inflationary minimum but also generates quantum corrections to the inflaton effective action. The most widely studied effect is the generation of radiative corrections to the scalar potential, which in SUSY models may be flat enough to sustain a sufficiently long period of slow-roll evolution if other SUSY breaking and supergravity corrections are appropriately suppressed. 

A less obvious but potentially more interesting effect arises from non-local corrections to the quantum effective action, which lead to dissipative effects. In an adiabatic regime typical of the slow-roll evolution, part of the inflaton's energy is then dissipated into other degrees of freedom, which may be the waterfall fields themselves or their decay products. If these degrees of freedom are relativistic and thermalize faster than expansion, this leads to the continuous sourcing of a radiation bath during inflation. This counteracts the dilution effect of Hubble expansion, eventually reaching a slowly evolving quasi-thermal state that may significantly modify the dynamics and predictions of inflation for temperatures $T>H$, where the Hubble rate $H$ gives the temperature of the quasi-de Sitter horizon. Hybrid constructions may thus naturally implement the idea of a warm rather than supercooled inflationary scenario, as originally proposed in \cite{Berera:1995wh, Berera:1995ie}.

Warm inflation has been successfully implemented in supersymmetric models, where the inflaton couples to fields that are unstable against decay into light particles and thus mediate dissipative effects \cite{Berera:2002sp, Moss:2006gt, BasteroGil:2009ec, BasteroGil:2010pb, BasteroGil:2012cm}. At sufficiently large inflaton field values, these mediator fields acquire masses that are large compared to the ambient temperature, so that their contribution to the inflaton effective potential is Boltzmann-suppressed and does not spoil the flatness of the scalar potential. Moreover,  supersymmetry suppresses the resulting radiative corrections, despite the finite temperature of the radiation bath \cite{Hall:2004zr}. From the discussion above, it is clear that SUSY hybrid inflation is a natural framework for this generic scenario, with the waterfall fields dissipating the inflaton's energy into light degrees of freedom that may include the Standard Model matter and gauge fields.

While the dynamics and perturbation spectra in warm hybrid inflation have been analyzed in \cite{BasteroGil:2009ec} (see also \cite{Ramos:2001zw} for related earlier work), as well as in \cite{BasteroGil:2011mr} for an analogous D-brane construction, using the approximate forms of the effective potential and dissipation coefficient for large field values, the evolution of the system close to the hybrid transition remains largely unexplored. One expects, in particular, that in this regime thermal corrections to the mass of the waterfall (super)fields may have a significant effect. These arise from the coupling of the waterfall fields to the light particles in the thermal bath and the resulting increase in the mass should sustain the metastable inflationary minimum parametrically below the critical inflaton field value at zero temperature.

In this work, we explore in detail the effects of thermal corrections to the effective potential, showing that they indeed generically delay the onset of the waterfall transition below the $T=0$ critical field value. We include in our analysis the full form of the finite temperature effective potential and dissipation coefficient, the complexity of which requires the use of numerical simulations.

We begin by introducing the SUSY hybrid inflation model in the next section, computing the finite temperature effective potential and the different contributions to the dissipation coefficient. In section 3, we review the dynamics of hybrid inflation in the supercooled regime ($T\ll H$) and in the warm regime ($T>H$) using the large field form of the potential and dissipation coefficient, where thermal corrections can be neglected. We show, in section 4, that a consistent inclusion of finite temperature corrections requires the use of an improved form of the effective potential, which we determine using the Cornwall-Jackiw-Tomboulis formalism for composite operators \cite{Cornwall:1974vz}. We use this result to numerically analyze the dynamics of warm hybrid inflation with finite temperature corrections in both the effective potential and dissipation coefficient, presenting and discussing our results in section 5. We summarize our main conclusions and discuss possible 
extensions of our analysis in section 6.


\section{Supersymmetric hybrid inflation}

We will consider in this work a model of SUSY hybrid inflation with chiral superfields $\Phi$, $X$ and $Y$ corresponding to the inflaton, waterfall and light sectors and described by a superpotential of the form  \cite{Moss:2006gt, BasteroGil:2009ec, BasteroGil:2010pb, BasteroGil:2012cm}:
\begin{equation} \label{superpotential}
W=g\Phi(X^2-M^2)+hXY^2~,
\end{equation}  
where $M$ is a constant mass parameter setting the scale of inflation and $g$ and $h$ are taken as real couplings (see \cite{BasteroGil:2011cx} for a generalization to the complex case that may lead to the generation of a baryon asymmetry during inflation). This simple superpotential will be sufficient to describe the main dynamical features of the model, although our results can be easily generalized to include charged waterfall superfields with trilinear couplings in the superpotential $g_{ij}\Phi X_i\bar{X}_j$, $i=1,\ldots, N_X$, with $X_i$ and $\bar{X}_j$ transforming in conjugate representations of a given gauge group. This is the case e.g.~of the D-brane construction considered in \cite{BasteroGil:2011mr} or, in the NMSSM, of the coupling between the Higgs chiral superfields and the singlet inflaton $g\Phi H_u H_d$ (see \cite{Ellwanger:2009dp} for a recent review). In this second example, the light $Y$ will also correspond to different quark and lepton chiral multiplets with Yukawa terms of the form in Eq.~(\ref{superpotential}), the same occurring for SUSY GUT extensions of the MSSM. For our purposes, it will be sufficient to consider $N_X$ species of waterfall supermultiplets and $N_Y$ light species with interactions given in terms of the same effective couplings $g$ and $h$. 

Denoting the classical scalar inflaton vacuum expectation value as $\phi=\sqrt{2}\langle\Phi\rangle$, which may be taken as real, and the scalar components in the waterfall and light sectors as $\chi$ and $y$, respectively, one obtains the following scalar interactions:
\begin{eqnarray} \label{scalar_lagrangian}
\mathcal{L}_{s}&=&2g^2\phi^2|\chi|^2+g^2|\chi^2-M^2|^2+\sqrt{2}hg\phi\left(\chi y^{\dagger 2}+\chi^\dagger y^2\right)+h^2|y|^4+4h^2|\chi|^2|y|^2~.
\end{eqnarray}  
Note that the scalar Lagrangian density may also include contributions to the inflaton potential from both supersymmetric and non-supersymmetric sources but since these are not crucial to our subsequent analysis we will omit them for simplicity. We also note that Eq.~(\ref{scalar_lagrangian}) is written for a single species in the $X$ and $Y$ sectors to simplify the notation, but we point out that scalar self-interactions in the these sectors are of the form $\chi_i^2\chi_j^{\dagger2}$ and $y_i^2 y_j^{\dagger 2}$, which will be relevant for our subsequent discussion, referring the reader to \cite{BasteroGil:2012cm} for further details in the more general case.

Similarly, the relevant interactions involving the fermionic components of the $X$ and $Y$ supermultiplets, $\psi_\chi$ and $\psi_y$, respectively, are given by:
\begin{eqnarray} \label{fermion_lagrangian}
\mathcal{L}_{fermion}&=&\sqrt2g\phi\bar\psi_\chi P_L\psi_\chi + 2h\chi\bar\psi_{y}P_L\psi_{y}+hy\bar\psi_{y}P_L\psi_\chi+\mathrm{h.c.}~,
\end{eqnarray}  
where again we have considered a single species in each sector and we have omitted interactions involving the inflaton superpartner, which will not affect our discussion.

From Eqs.~(\ref{scalar_lagrangian}) and (\ref{fermion_lagrangian}) we easily deduce that, while the fields in the $Y$ sector remain massless (up to additional tree-level contributions that we assume to be negligible), the fields in the waterfall sector acquire masses:
\begin{eqnarray} \label{waterfall_masses}
m_{\chi_R}^2&=&2g^2(\phi^2-M^2) ~,\\\nonumber
m_{\chi_I}^2&=&2g^2(\phi^2+M^2)~,\\\nonumber
m_{\psi_\chi}^2&=&2g^2\phi^2~,
\end{eqnarray}
with $\chi=(\chi_R+i\chi_I)/\sqrt2$, keeping $\mathrm{Str}M_X^2= m_{\chi_R}^2+m_{\chi_I}^2-2m_{\psi_\chi}^2=0$ as typical of F-term SUSY breaking models. This shows explicitly that for $\phi>M$ the scalar potential for the $\chi$ scalar waterfall fields has a non-SUSY minimum at the origin with potential energy $V_0=g^2M^4$, while for $\phi<M$ the ground state is supersymmetric at $\chi=M$. Hence, $\phi_c=M$ gives the critical value of the inflaton field at which the second order waterfall phase transition takes place in the supercooled inflationary scenario.

The bosonic and fermionic components in the $X$ sector will then contribute to the inflaton effective potential via the standard Coleman-Weinberg contribution at 1-loop \cite{Coleman:1973jx}, yielding for the full inflaton potential:
\begin{eqnarray} \label{Coleman-Weinberg}
V(\phi)=V_0 +{N_X\over 64\pi^2}\sum_{i=\chi_{R,I}, \psi_\chi}m_i^4\left[\log\left(m_i^2\over \mu^2\right)-{3\over 2}\right]~,
\end{eqnarray}
where $\mu$ is the renormalization scale. Note that since the mass of the $Y$ sector fields is independent of $\phi$ these do not contribute to the expression above. In the large field limit, $\phi\gg M$, the potential takes the simple form:
\begin{eqnarray} \label{potential_cold_large}
V(\phi)=V_0\left[1+\gamma\log\left({\phi\over \mu}\right)\right]~,
\end{eqnarray}
with $\gamma=g^2N_X/(4\pi^2)$ controlling the size of radiative corrections.

As mentioned above, the coupling between the inflaton and waterfall fields, namely the first terms in Eqs.~(\ref{scalar_lagrangian}) and (\ref{fermion_lagrangian}), lead to non-local terms in the effective action, which in the adiabatic approximation give an additional friction term of the form $\Upsilon\dot\phi$ in the inflaton's equation of motion. It is clear from the model above that the fields in the $X$ sector are unstable against decay into the light $Y$ supermultiplets, in particular with $\chi\rightarrow yy, \psi_y\psi_y$ and $\psi_\chi\rightarrow y \psi_y$. Dissipative interactions will then transfer part of the inflaton's energy into the $Y$ sector, which may thermalize via decay, inverse decays and scattering processes forming a radiation bath concurrent with accelerated expansion. The leading contributions to the dissipative coefficient $\Upsilon$ have been computed in \cite{BasteroGil:2010pb, BasteroGil:2012cm} in the low temperature regime, where $m_{\chi_{R,I}}, m_{\psi_\chi}>T$, with the fermionic contribution being subdominant. These were computed neglecting the SUSY mass splittings within the $X$ chiral multiplets, which holds for large field values according to Eq.~(\ref{waterfall_masses}), but it is easy to extend this result for generic scalar masses, yielding:
\begin{eqnarray} \label{dissipation_coefficient}
\Upsilon&=&\Upsilon_{LM}+\Upsilon_P\nonumber\\
&=&\sum_{i=\chi_{R,I}}\left[0.64h^{2}g^8N_XN_Y{T^3\phi^6\over m_i^8}+{16\over\sqrt{2\pi}}{g^2N_X\over h^2N_Y}\left({2g^2\phi^2\over 2g^2\phi^2+m_i^2}\right)\sqrt{Tm_i}e^{-m_i/T}\right]~,
\end{eqnarray}
where the first and second terms give the contributions from virtual low-momentum modes  (``LM") and from on-shell modes corresponding to poles in the scalar propagator (``P"), respectively. For low-momentum $\chi_{R,I}$ modes, the dominant decay channel is $\chi_{R,I}\rightarrow yy$, while for on-shell modes both scalar and fermionic decay channels may contribute, with total decay width:
\begin{eqnarray} \label{decay_width}
\Gamma_{\chi_i}= \Gamma(\chi_i\rightarrow yy)+\Gamma(\chi_i\rightarrow \psi_y\psi_y)= {h^2 N_Y\over 16\pi}\left({2g^2\phi^2\over m_i}+m_i\right)~.
\end{eqnarray}
Note that this corresponds to the on-shell zero-temperature result for $\mathbf{p}=0$, which as shown in \cite{BasteroGil:2012cm} gives a sufficiently good approximation to the full on-shell contribution to the dissipation coefficient. On the other hand, thermal effects enhance the width in the low-momentum regime, which is accounted for in the expression above. Also notice that, for $\phi\gg M$, both decay channels occur with equal probability in the on-shell regime.

The contributions from low-momentum and on-shell modes are dominant in different regimes. On one hand, on-shell production leads to a resonant enhancement of the dissipation coefficient but, on the other hand, the corresponding occupation numbers are Boltzmann-suppressed in the non-relativistic regime. Generically, low-momentum modes give the leading contribution to dissipation for large values of $m_{i}/T$, since they are not exponentially suppressed, but for $m_i/T\gtrsim 1$ on-shell modes become the dominant mediators of dissipative processes.

Hence, in the large field limit, the second term in Eq.~(\ref{dissipation_coefficient}) is negligible and the low-momentum contribution takes the simple form:
\begin{eqnarray} \label{dissipation_coefficient_LM}
\Upsilon_{LM}=C_\phi {T^3\over \phi^2}~, \qquad C_\phi=0.08 h^2N_XN_Y~.
\end{eqnarray}
We refer the reader to \cite{BasteroGil:2010pb, BasteroGil:2012cm} for the details of this computation, noting that it assumes also the conditions $T>H$, where a flat space computation is valid, and $\Gamma_\chi>H, \dot\phi/\phi$, which ensures an adiabatic evolution of the field and expansion compared to the relevant microphysical processes and that near-thermal equilibrium configurations can be maintained during inflation (see e.g. \cite{BasteroGil:2009ec}). Also, the perturbative nature of the calculation requires that, for finite inflaton values, $h^2N_Y\lesssim 1$.

Since dissipation can sustain a nearly-thermal bath of radiation with a slowly varying temperature during inflation, we must also take into account finite temperature corrections to the effective potential, generically given by \cite{Dolan:1973qd}:
\begin{eqnarray} \label{finite-temperature-potential}
\Delta V_i^{(T)}={T^4\over 2\pi^2}\mathrm{Str}\int_0^\infty dxx^2\log\left(1\mp e^{-\sqrt{x^2+m_i^2/T^2}}\right)~,
\end{eqnarray}
for all fields in the $X$, $Y$ and inflaton sectors. On one hand, it is well known that this gives an overall leading contribution $-(\pi^2/90)g_*T^4$ from all relativistic fields, with $g_*=g_b+(7/8)g_f$ denoting the effective number of bosonic and fermionic relativistic degrees of freedom (see e.g. \cite{Cline:1996mga}). For non-relativistic fields, on the other hand, this contribution is Boltzmann-suppressed and the zero-temperature Coleman-Weinberg potential is the leading correction. 

To determine the regimes in which the different sectors contribute as relativistic or non-relativistic fields to the effective potential, one must also take into account thermal corrections to two-point correlation functions, which in particular change the field masses. It is, in fact, well known that a resummation of the so-called daisy and superdaisy diagrams results in the inclusion of thermal mass corrections to field propagators, so that full thermal masses should be used in both the Coleman-Weinberg contribution, which is otherwise independent of the temperature, and the finite temperature contribution in Eq.~(\ref{finite-temperature-potential}). One must note that in scalar theories with quartic self-interactions this procedure leads, however, to a double counting of the `figure of eight' diagram \cite{Dine:1992wr, Espinosa:1992gq}, an issue we will return to in section 4.

Thermal corrections to the masses in the different sectors of the warm hybrid construction have been discussed in detail in \cite{Hall:2004zr} and here we will only summarize the main results. As mentioned above, we are mainly interested in the low-temperature regime, where the fields in the $X$ sector are heavy and hence thermal corrections to the inflaton mass are Boltzmann-suppressed. In this regime, the leading $X$ sector contribution to the effective potential is thus the Coleman-Weinberg potential in Eq.~(\ref{Coleman-Weinberg}). However, since they are coupled via bi-quadratic and Yukawa interactions to the fields in the $Y$ sector, which are massless at tree-level as seen above, they receive thermal mass corrections which, due to the underlying supersymmetry, are equal for all bosonic and fermionic degrees of freedom:
\begin{eqnarray} \label{waterfall_masses_temperature}
m_{\chi_R}^2&=&2g^2(\phi^2-M^2)+\alpha^2T^2 ~,\\\nonumber
m_{\chi_I}^2&=&2g^2(\phi^2+M^2)+\alpha^2T^2~,\\\nonumber
m_{\psi_\chi}^2&=&2g^2\phi^2+\alpha^2T^2~,
\end{eqnarray}
where $\alpha^2=h^2N_Y/2$ \cite{BasteroGil:2012cm}. The fields in the $Y$ sector also acquire a thermal mass due to their self-interactions but, as shown in  \cite{BasteroGil:2012cm} , $m_{y,\psi_y}^2\sim \mathcal{O}(h^2)$, so that these may be neglected for $h\ll 1$ and a radiation bath with $g_*=(15/4)N_Y$ is sourced during inflation. This also justifies the Hard Thermal Loop (HTL) approximation used in computing the thermal corrections to the $X$ sector masses above. Since in the general case the waterfall $X$ sector fields, as well as the $Y$ fields, may be charged under the Standard Model gauge group or a particular extension, gauge bosons may also contribute to the waterfall thermal masses, with e.g. $\Delta m_X^2\sim \mathcal{O}(g_{YM}^2 N_c)$ for a $SU(N_c)$ gauge group with Yang-Mills coupling $g_{YM}$, as well as contributing to the number of relativistic degrees of freedom. To account for this and other more general possibilities, we will treat both $\alpha$ and $g_*$ as generic parameters in  our analysis, keeping in mind the values given above for the case where the thermal bath is made only of the relativistic $Y$ multiplets. The inflaton field and its superpartners will either behave effectively as additional relativistic or non-relativistic species, so that for large field multiplicities they will give a sub-dominant contribution to the effective potential.

Taking all thermal corrections into account, we thus arrive at the following leading form for the effective potential at finite temperature:
\begin{eqnarray} \label{potential_xi}
V(\phi,T)&=&V_0\left\{1+{\gamma\over16}\left[\xi^2\left(\log\left({\xi\over \bar\mu^2}\right)-{3\over2}\right)+(\xi+4)^2\left(\log\left({\xi+4\over \bar\mu^2}\right)-{3\over2}\right)\right.\right.\nonumber\\
&-&\left.\left.2(\xi+2)^2\left(\log\left({\xi+2\over \bar\mu^2}\right)-{3\over2}\right)\right]\right\}
-{\pi^2\over 90}g_*T^4~,
\end{eqnarray}
where we have defined the normalized mass squared of the scalar waterfall fields:
\begin{eqnarray} \label{xi}
\xi(\phi,T)={m_{\chi_R}^2(\phi,T)\over g^2M^2}={2g^2(\phi^2-M^2)+\alpha^2T^2\over g^2 M^2}~,
\end{eqnarray}
such that at finite temperature the inflationary minimum becomes unstable when $\xi(\phi,T)<0$. We have also defined the dimensionless renormalization scale $\bar\mu=\mu/gM$.

This form of the effective potential and the dissipation coefficient in Eq.~(\ref{dissipation_coefficient}), both valid in the low-temperature regime $m_{\chi_{RI},\psi_\chi}\gtrsim T$ and including thermal corrections to the field masses, provide the necessary ingredients to analyze the dynamics of warm hybrid inflation. 
  

\section{Preliminary analysis: inflationary dynamics at large field values}

When the effects of dissipative interactions with other fields are taken into account, the equation of motion for the scalar inflaton field takes the form (see e.g. \cite{Berera:2008ar} for a review):
\begin{eqnarray} \label{inflaton_equation}
\ddot{\phi}+3H\dot\phi+\Upsilon\dot\phi+V_\phi=0~,
\end{eqnarray}
where $V_\phi$ denotes the first derivative of the scalar potential with respect to the inflaton field. When dissipation results in the production of relativistic degrees of freedom, as occurs in the hybrid model that we are considering, the total energy density and pressure of the system are given by:
\begin{eqnarray} \label{total_rho_p}
\rho&=&{1\over2}\dot\phi^2+V(\phi,T)+Ts~,\nonumber\\
p&=& {1\over2}\dot\phi^2-V(\phi,T)~,
\end{eqnarray}
where $s=-\partial V(\phi,T)/\partial T\equiv - V_T$ is the entropy density of the system, which is well defined when the system remains close to equilibrium throughout inflation. The covariant conservation of the total energy-momentum tensor follows the standard continuity equation in an expanding universe, $\dot\rho+3H(\rho+p)=0$, from which we can deduce, upon using Eq.~(\ref{inflaton_equation}):
\begin{eqnarray} \label{entropy_equation}
\dot{s}+3Hs={\Upsilon\dot\phi^2\over T}~.
\end{eqnarray}
When thermal corrections to the field masses have a negligible effect, which from the discussion above corresponds to the large field regime, $\phi\gg M, (\alpha/\sqrt2g)T$, the effective potential decouples into a field and a temperature dependent part, $V(\phi,T)=V(\phi)-(\pi^2/90)g_* T^4$, with the standard expression for the entropy density of a gas of relativistic particles, $s=(2\pi^2/45)g_*T^4$. The corresponding energy density is thus $\rho_R= Ts-(\pi^2/90)g_* T^4= (\pi^2/30)g_*T^4$, and from Eq.~(\ref{entropy_equation}) we get:
\begin{eqnarray} \label{radiation_equation}
\dot{\rho_R}+4H\rho_R=\Upsilon\dot\phi^2~.
\end{eqnarray}
This separation is not straightforward in the general case, as we will analyze in the next section, but we will first focus on this simplified regime where the main features of warm inflation are easily described. In particular, it is easy to see that dissipation indeed acts as a source term that counteracts the effects of expansion, which would otherwise dilute any primordial radiation component exponentially, leading to a supercooled inflationary universe. In warm inflation, on the other hand, the system quickly reaches a slowly-evolving configuration with $\dot\rho_R\simeq 0$, with the slow-roll equations yielding:
\begin{eqnarray} \label{slow_roll_eqs}
3H(1+Q)\dot\phi&\simeq& -V_\phi~, \nonumber\\
4H\rho_R&\simeq &\Upsilon \dot{\phi}^2~,
\end{eqnarray}
where $Q\equiv \Upsilon/3H$, from which we can determine both the evolution of both the field and the temperature of the radiation during inflation. A slow-roll evolution is obtained in this case when the following conditions are satisfied:
\begin{eqnarray} \label{slow_roll_conditions}
\epsilon_\phi={M_P^2\over 2}\left(V_\phi\over V\right)^2\ll 1+Q~, \qquad |\eta_\phi|=M_P^2{|V_{\phi\phi}|\over V}\ll 1+Q~,
\end{eqnarray}
which generalize the standard slow-roll conditions in the supercooled case, showing that warm inflation allows for steeper scalar potentials than the latter if dissipation becomes sufficiently strong at some stage during inflation. Accelerated expansion requires also the radiation energy density to be subdominant, yielding the Friedmann equation $H^2\simeq V/3M_P^2$, and from Eqs.~(\ref{slow_roll_eqs}) we can deduce that:
\begin{eqnarray} \label{radiation_abundance}
{\rho_R\over V}\simeq {\epsilon_\phi\over2}{Q\over (1+Q)^2}~,
\end{eqnarray}
so that radiation remains a sub-leading component in the slow-roll regime. The radiation abundance may nevertheless increase if dissipation becomes strong, $Q\gg 1$, such that at the end of the slow-roll regime, for $\epsilon_\phi\simeq 1+Q$, it already yields a significant fraction of the energy density. In this case, it will typically come to dominate quickly afterwards, providing a smooth exit into the Standard Hot Big Bang cosmology with no need for a separate reheating period.


\subsection{Cold regime}

Let us first review the dynamics of SUSY hybrid inflation in the cold regime \cite{Dvali:1994ms}, $T\ll H$, in the large field limit where the scalar potential takes the logarithmic form in Eq.~(\ref{potential_cold_large}), assuming that radiative corrections from the coupling to the waterfall field(s) are the dominant effect. Dissipation has, in principle, a negligible effect in this case, although to our knowledge no analysis of dissipative effects for $T\ll H$ has been performed to date. Assuming also that radiative corrections are subdominant compared to the constant vacuum energy $V_0$, the slow-roll parameters take the approximate form:
\begin{eqnarray} \label{SR_parameters}
\epsilon_\phi\simeq{\gamma^2\over2}\left({M_P\over \phi}\right)^2~,\qquad
\eta_\phi\simeq-\gamma\left({M_P\over \phi}\right)^2~,
\end{eqnarray}
with inflation ending for $\phi_e\simeq \gamma/\sqrt2$. It is also easy to compute the number of e-folds of inflation from the moment that a given pivot CMB scale leaves the horizon during inflation, at $\phi=\phi_*\gg \phi_e$, giving:
\begin{eqnarray} \label{e_folds_cold}
N_e\simeq -M_P^{-2}\int_{\phi_*}^{\phi_e}{V\over V_\phi} d\phi \simeq {1\over2\gamma}{\phi_*^2\over M_P^2}~.
\end{eqnarray}
While the amplitude of the resulting density perturbation spectrum can be easily obtained by choosing the constant $V_0$, the corresponding spectral index and tensor-to-scalar ratio are given, for small $\gamma$, by:
\begin{eqnarray} \label{cold_observables}
n_s-1=2\eta_{\phi_*}-6\epsilon_{\phi_*}\simeq -{1\over N_e}~,\qquad 
r=16\epsilon_{\phi_*}\simeq 4\gamma (1-n_s)~.
\end{eqnarray}
This yields $n_s>0.98$ for $N_e>50$, which has been excluded by the observations of the Planck satellite at 95\% C.L., although the tensor-to-scalar ratio can be within Planck's upper bound $r<0.11$ for $\gamma<0.68$ if the observed value of $n_s$ is assumed \cite{Ade:2013uln}.


\subsection{Warm regime}

In the warm case, $T\gtrsim H$, we may use the large field form of the dissipation coefficient in Eq.~(\ref{dissipation_coefficient_LM}), which corresponds to mediation by virtual low-momentum waterfall field modes. Using $\Upsilon=C_\phi T^3/\phi^2$ in the slow-roll equations, we can derive the following general relations \cite{BasteroGil:2009ec}:
\begin{eqnarray} \label{warm_relations}
{T\over H}&=&\left({3\over2C_R}\right)^{1/4}{Q^{1/4}\over (1+Q)^{1/2}}\left(V(\phi)\over 3M_P^4\right)^{-1/4}\epsilon_\phi^{1/4}~,\nonumber\\
Q^{1/3}(1+Q)^2&=&2\epsilon_\phi\left({C_\phi\over 3}\right)^{1/3}\left({C_\phi\over 4C_R}\right)\left({H\over M_P}\right)^{2/3}\left({\phi\over M_P}\right)^{-8/3}~,
\end{eqnarray}
and use them to express both the field and the temperature as a function of the dissipative ratio $Q$, which then satisfies the following differential equation :
\begin{eqnarray} \label{Q_eq_hybrid}
{Q'\over Q}\simeq C_Q {Q^{1/7}(1+Q)^{6/7}\over 1+7Q}~,
\end{eqnarray}
with $C_Q=14\gamma (M_P/ \phi_*)^2Q_*^{-1/7}(1+Q_*)^{-6/7}$, $Q_*$ denoting the dissipative ratio at horizon-crossing for the chosen pivot scale and primes denoting derivatives with respect to the number of e-folds of inflation, $dN_e= Hdt$. This shows that dissipation becomes stronger as inflation proceeds and, considering a regime where dissipation is weak at horizon-crossing, $Q_*\ll1$, we obtain the following approximate expression for the number of e-folds of inflation in the warm regime:
\begin{eqnarray} \label{e_folds_warm}
N_e\simeq  {1\over2\gamma}{\phi_*^2\over M_P^2}\left[1+(\log Q_e-b)Q_*^{1/7}\right]~,
\end{eqnarray}
where $b=1+\gamma_E+\psi(0,6/7)\simeq 0.74$. Comparing with Eq.~(\ref{e_folds_cold}), we thus see explicitly that dissipation enhances the number of e-folds of inflation if $Q_e\gg 1$. The slow-roll conditions are only violated in this case when radiative corrections become significant, which occurs close to the hybrid transition. If thermal corrections to the waterfall field masses are neglected, this occurs for $\phi_e\simeq M$ and $Q_e\simeq (\phi_*/M)^2Q_*^{1/7}$, which is typically large if dissipation is not too weak at horizon-crossing.

The spectrum of density perturbations in warm inflation is modified by fluctuation-dissipation effects \cite{Berera:1995wh, Berera:1995ie, Berera:1999ws, Hall:2003zp} and possibly by the fact that inflaton perturbations may be in a non-trivial statistical state \cite{Ramos:2013nsa}, which occurs if interactions between the inflaton and light degrees of freedom in the thermal bath occur sufficiently fast \cite{Bartrum:2013fia}. Additionally, perturbations in the field and radiation fluid are coupled through the temperature dependence of the dissipation coefficient \cite{Graham:2009bf}, an effect which may be significant if dissipation is strong at horizon-crossing. In the simplest scenario, where no large inflaton occupation numbers are generated during inflation and $Q_*\ll1$, the scalar spectrum takes the approximate form \cite{Bartrum:2013fia}:
\begin{eqnarray} \label{warm_scalar_spectrum}
\Delta_\mathcal{R}^2={1\over24\pi^2}{V(\phi_*)\over M_P^4}\epsilon_{\phi_*}^{-1}\left(1+\kappa_*\right)~,
\end{eqnarray}
where $\kappa_*=2\pi Q_*T_*/H_*$ parametrizes the leading modification to the cold inflaton expression for $Q_*\ll 1$ and $T_*\gtrsim H_*$. Since tensor modes only couple gravitationally to the thermal bath and hence are not significantly modified by fluctuation-dissipation dynamics, we obtain the following expressions for the spectral index and tensor-to-scalar ratio:
\begin{eqnarray} \label{warm_observables}
n_s-1&\simeq&2\eta_{\phi_*}-6\epsilon_{\phi_*}+{\kappa_*\over1+\kappa_*}\left(14\epsilon_{\phi_*}-8\eta_{\phi_*}+10\sigma_{\phi_*}\right)\simeq -2\gamma\left({M_P\over \phi_*}\right)^2\left({1-8\kappa_*\over1+\kappa_*}\right)~,\nonumber\\
r&\simeq& {16\epsilon_{\phi_*}\over 1+\kappa_*}\simeq {4\gamma(1-n_s)\over 1-8\kappa_*} ~,
\end{eqnarray}
where $\sigma_\phi=M_P^2 V_\phi/(V\phi)$. Note that these reduce to Eq.~(\ref{cold_observables}) for $\kappa_*\ll1$, so that in the warm regime one can obtain a similar spectrum of perturbations but with a larger number of e-folds. On the other hand, if $\kappa_*>1/8$ the spectrum becomes blue-tilted, which has been ruled out by Planck, thus favouring a scenario where dissipation is still weak when observable scales leave the horizon during inflation, $Q_*\ll1$, as assumed above. In Figure 1 we show a numerical solution of the slow-roll equations taking $\Upsilon=C_\phi T^3/\phi^2$ and neglecting thermal mass corrections in the 1-loop Coleman-Weinberg potential for a particular choice of parameters yielding $n_s\simeq 0.962$, $r\simeq 0.019$ and 60 e-folds of inflation after horizon-crossing, within Planck's observational window \cite{Ade:2013uln}.

\begin{figure}[htbp] \label{observables}
\centering\includegraphics[scale=1.2]{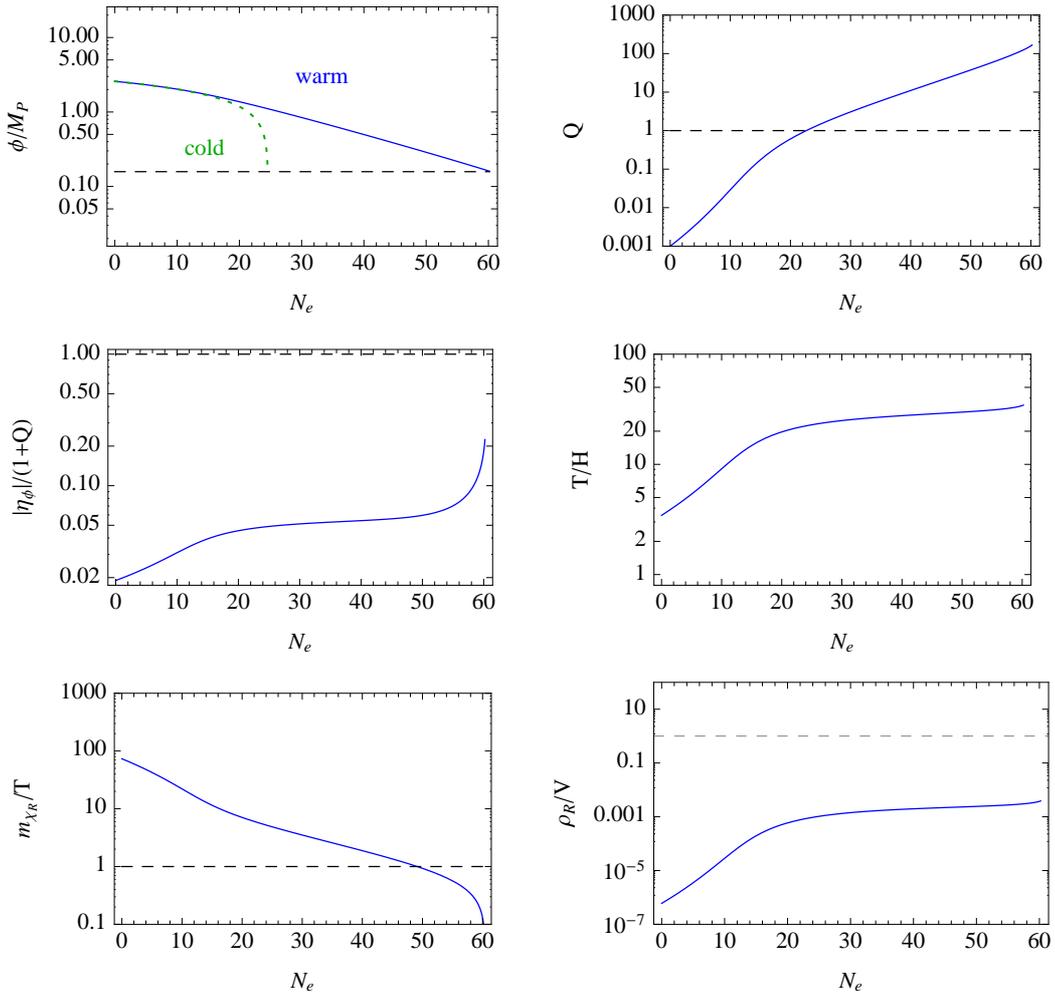}
{\vskip-0.3cm}
\caption{Numerical evolution of the different dynamical quantities in the warm regime (solid blue curves), using the slow-roll equations for $\Upsilon=C_\phi T^3/\phi^2$ and neglecting thermal corrections to the potential. For this example, $g=10^{-3}$, $h=0.34$, $N_X=5\times10^6$, $N_Y=50$ and $M=0.16 M_P$, choosing $Q_*\simeq 10^{-3}$. For this choice of parameters, the spectral index is $n_{s}\simeq0.962$ and the tensor-to-scalar ratio $r=0.019$, with the amplitude of the power spectrum normalized to the observational value $\Delta_\mathcal{R}^{2}\approx2.2\times10^{-9}$. In the top left plot we also show the corresponding numerical solution in the supercooled regime (dashed green curve) and the the value of the mass scale $M$ (black dashed line).}
\end{figure}

As one can see in this example, the system evolves deeper into the warm regime, i.e. the ratio $T/H$ increases, as inflation proceeds, and dissipation becomes stronger, thus more than doubling the number of e-folds of inflation with respect to the supercooled case. The relative abundance of radiation also increases during inflation. By the time the field reaches the critical value $\phi_c=M$, it already accounts for almost 1\% of the total energy density, and it should quickly take over as the system flows into the supersymmetric ground state. One must note that the waterfall fields become relativistic about 10 e-folds before the end of inflation, so that the low-temperature approximation is only valid for about 50 e-folds of inflation in this example. However, we will see in the next section that the inclusion of thermal corrections to the effective potential and the contribution of on-shell modes to dissipation may become significant before the hybrid transition, modifying the number of e-folds in the low-temperature approximation. 

In this example, a large number of waterfall fields was considered, which is generically required in order to keep the system in the warm regime with the form of the dissipation coefficient considered, as observed in \cite{BasteroGil:2009ec, Cerezo:2012ub}. Such large field multiplicities may be obtained in multiple D-brane implementations of SUSY hybrid inflation \cite{BasteroGil:2011mr} or potentially in extra-dimensional models \cite{Matsuda:2012kc}. 

It is also worth mentioning that if a thermal distribution of inflaton particles is present at horizon-crossing and can be sustained during the remainder of inflation, as discussed in  \cite{Bartrum:2013fia}, the spectrum of scalar perturbations is considerably modified, yielding in particular a spectral index:
\begin{eqnarray} \label{n_s_thermal}
n_s-1\simeq 2\sigma_{\phi_*}-2\epsilon_{\phi_*}\simeq \gamma\left(2-\gamma\right)\left(M_P\over \phi_*\right)^2~.
\end{eqnarray}
This yields a blue-tilted spectrum, which has been ruled out by Planck, unless $\gamma>2$, i.e. $g^2N_X>8\pi^2$. In this parametric regime, higher-order corrections to the effective potential may become important and must be carefully analyzed, which is however outside the scope of this work. We nevertheless note that the inflaton particle production rate through the 3-body decay $\chi \rightarrow \phi yy$ becomes more significant for larger values of $\gamma$, as discussed in  \cite{Bartrum:2013fia}, which suggests that it may indeed be appropriate to consider a thermal inflaton state for large $\gamma$ and a quasi-vacuum state for small $\gamma$, potentially yielding consistent observational predictions in both regimes.


\section{Finite temperature corrections in the CJT formalism}

As discussed earlier, the masses of the waterfall fields and their superpartners receive thermal corrections from their coupling to the thermal bath of relativistic fields, yielding an effective potential of the form (\ref{potential_xi}). This would imply that, even though the $X$ sector fields are non-relativistic and have Boltzmann-suppressed occupation numbers, they may contribute to the entropy density of the system through the temperature dependence of their effective masses:
\begin{eqnarray} \label{X_entropy}
s_X=-\sum_i {\partial V\over \partial m_i^2}{\partial m_i^2\over \partial T}~,
\end{eqnarray}
where the index $i$ is summed over all components of the $N_X$ waterfall supermultiplets. This yields, in particular:
\begin{eqnarray} \label{entropy_heavy}
s_X=-{\gamma\over 4} \alpha^2 M^2 T F(\xi)~, 
\end{eqnarray}
where $\xi$ is the normalized waterfall mass defined in Eq.~(\ref{xi}) and
\begin{eqnarray} \label{entropy_function}
F(\xi)=\xi\log(\xi)+(\xi+4)\log(\xi+4)-2(\xi+2)\log(\xi+2)~.
\end{eqnarray}
Firstly, note that this is independent of the renormalization scale, although the effective Coleman-Weinberg potential is not, which is simply a consequence of the supertrace condition $\mathrm{Str}\mathcal{M}_X^2=0$ in the waterfall sector. Secondly, it is easy to check that this function is positive-definite for $\xi>0$, i.e. in the inflationary region, with $0<F(\xi)<4\log(2)$, so that $s_X<0$. This suggests that the waterfall fields could reduce the total entropy density of the system, which includes the standard contribution from relativistic degrees of freedom. However, since it scales linearly with the temperature, $s_X$ may actually overcome the relativistic contribution $\propto T^3$ at low temperatures to yield a {\it negative} total entropy. For example, at large field values we find:
\begin{eqnarray} \label{entropy_large_field}
s\simeq {2\pi^2\over 45}g_* T^3\left(1-{3\over4}{\gamma V_0\over \rho_R}{\alpha^2T^2\over m_X^2}\right)~,
\end{eqnarray}
so that even if thermal mass corrections are subdominant in this regime, $\alpha T\ll m_X\simeq \sqrt{2}g\phi$, the total entropy density may become negative if $\gamma V_0\ll \rho_R\ll V_0$, which is compatible with accelerated expansion for $\gamma\ll 1$.

Such a negative entropy is physically unacceptable, showing that the direct inclusion of thermal mass corrections in the Coleman-Weinberg potential is ill-defined. As discussed in section 2, the inclusion of thermal masses in field propagators is known to lead to inconsistencies, in particular to over-counting certain loop diagrams in the corresponding resummation procedure. An improved version of the effective potential can be obtained using the Cornwall-Jackiw-Tomboulis (CJT) formalism for composite operators \cite{Cornwall:1974vz}, a non-perturbative scheme that has been used in the description of several finite temperature quantum systems  \cite{AmelinoCamelia:1992nc, AmelinoCamelia:1996hw, Blaizot:2000fc, Blaizot:2006tk} and where a self-consistent expansion of the effective potential in terms of full propagators can be obtained. 

For simplicity, let us discuss the basic features of this formalism considering a real scalar field $\chi$ of bare mass $m$ and quartic self-interactions $(\lambda/4!)\varphi^4$ coupled to a thermal bath at temperature $T$. The effective potential or free-energy of the field is then given by:
\begin{eqnarray} \label{potential_CJT}
V=V^{(0)}-{i\over 2}\int_p \log G^{-1}-{1\over2}\int_p~\Pi G+\Phi[G]~,
\end{eqnarray}
where $V^{(0)}$ denotes the tree-level potential, $G$ is the full scalar propagator and $\Pi$ is the corresponding self-energy, such that:
\begin{eqnarray} \label{scalar_propagator}
G(p)={i\over p^2-m^2-\Pi}~, 
\end{eqnarray}
which should be distinguished from the {\it bare} propagator $G_0(p)=i/(p^2-m^2)$. We have also used the following short-hand notation for the momentum-space integrals, including the sum over the Matsubara frequencies in the imaginary-time formalism:
\begin{eqnarray} \label{momentum_integral}
\int_p f(p)=iT\sum_n \int {d^3p\over (2\pi)^3}f(i\omega_n, \mathbf{p})~.
\end{eqnarray}
The last term in Eq.~(\ref{potential_CJT}), $\Phi[G]$, corresponds to a sum of 2-particle irreducible diagrams, which corrects for the double-counting introduced when replacing the bare propagators with the full propagators in the effective potential. The CJT formalism leads to a self-consistent expansion of the effective potential by requiring the stationarity of the effective potential with respect to variations of the full propagator, for fixed $G_0$, such that:
\begin{eqnarray} \label{stationarity}
{\delta V\over \delta G}=0~.
\end{eqnarray}
This defines the self-energy and hence the physical propagator of the field in a consistent way through the gap equation:
\begin{eqnarray} \label{gap_equation}
{\delta\Phi \over \delta G}={1\over 2}\Pi~.
\end{eqnarray}
The gap equation expresses nothing more than the fact that the diagrams contributing to the self-energy are obtained by cutting an internal line in the corresponding 2-particle-irreducible diagrams included in $\Phi$ \cite{Blaizot:2000fc}.

For the case of a scalar with quartic self-interactions, the leading contributions to $\Phi$ arise at two-loop order and correspond to the `double-bubble' (or `figure of eight') and `sunset' diagrams illustrated below, of order $\lambda$ and $\lambda^2$, respectively. 

\vspace{0.5cm}
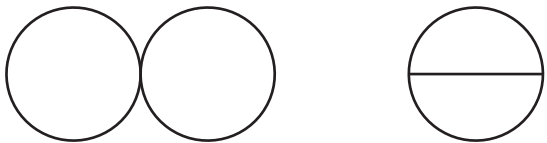
\begin{figure}[htbp]
\centering\fcolorbox{white}{white}{
  \begin{picture}(313,50) (40,-40)
    \SetWidth{1.0}
    \SetColor{Black}
    \Arc(120,0)(25,54,420)
    \Arc(170,0)(25,54,420)
     \Arc(270,0)(25,54,420)
     \Line(245,0)(295,0)
  \end{picture}
}
\caption{Two-loop diagrams contributing to the function $\Phi[G]$, including the ``double-bubble" or ``figure-of-eight" diagram (left) and the ``sunset" diagram (right).}
\end{figure}

Neglecting the latter diagram is thus typically a sufficiently good approximation, known as the `Hartree-Fock' or simply `double-bubble' approximation, which moreover simplifies the analysis considerably since the sunset diagram introduces a momentum-dependent term in the gap equation \cite{Nachbagauer:1994ur, Nemoto:1999qf} (see e.g. \cite{Baacke:2002pi, Verschelde:2000ta} for similar approaches including the sunset diagram). For the case of the real and imaginary waterfall components, the nature of the interactions implies that the double-bubble diagram is $\mathcal{O}(g^2)$ while the sunset diagram is $\mathcal{O}(g^4N_X)$, noting that there are no double-bubble diagrams involving different species, although there are `mixed' double-bubble diagrams involving the real and imaginary components of {\it each} chiral supermultiplet. The sunset diagram is thus clearly sub-leading in our supersymmetric hybrid inflation model even for $g^2N_X\lesssim 1$ if $g\ll 1$. Moreover, since during inflation $\langle \chi \rangle=0$, there is no trilinear vertex involving only the $\chi$ fields and the associated sunset diagram vanishes.

The effective potential for the generic scalar $\varphi$ can then be written as:
\begin{eqnarray} \label{CJT_potential_scalar}
V(\varphi)&=&{\lambda\over 4!}\varphi^4+{1\over2}\int_E\log(p^2+\Omega^2)-\nonumber\\
&-&{1\over2}P_T(\Omega)\left(\Omega^2-m^2-{\lambda\over 2}\varphi^2-\alpha^2T^2-{\lambda\over 4} P_T(\Omega)\right)~,
\end{eqnarray}
where $\Omega$ is the effective mass of the scalar and, as before, $\alpha$ parametrizes the generic thermal mass corrections from the coupling to the thermal bath. Note that the second term yields, after regularization, the standard Coleman-Weinberg potential and the finite temperature term in Eqs.~(\ref{Coleman-Weinberg}) and (\ref{finite-temperature-potential}), respectively. Following \cite{AmelinoCamelia:1996hw}, we have also defined:
\begin{eqnarray} \label{bubble}
P_T(\Omega)&=&\left.\int_p{i\over p^2-\Omega^2}\right|_{\overline{DR}}={\Omega^2\over 16\pi^2}\left(\log{\Omega^2\over \mu^2}-1\right)+\int{d^3p\over (2\pi)^3}{n_B(E)\over E}~,
\end{eqnarray}
which corresponds to the single $\varphi$-bubble. The stationarity condition corresponds in practice to variations of the CJT effective potential with respect to the physical scalar mass $\Omega$ for a constant bare mass, yielding:
\begin{eqnarray} \label{stationarity_scalar}
{\partial V\over \partial \Omega^2}=-{1\over2}{\partial P_T\over \partial \Omega^2}\left(M^2-m^2-{\lambda\over 2}\varphi^2-\alpha^2T^2-{\lambda\over2}P_T(\Omega^2)\right)=0~,
\end{eqnarray}
where we have used the identity:
\begin{eqnarray} \label{bubble_identity}
P_T(\Omega)={\partial\over\partial \Omega^2}{1\over2}\int_E\log(p^2+\Omega^2)~.
\end{eqnarray}
The gap equation determining the physical mass of the scalar field then reads:
\begin{eqnarray} \label{gap_equation_scalar}
\Omega^2=m^2+{\lambda\over 2}\varphi^2+\alpha^2T^2+{\lambda\over2}P_T(\Omega^2)~.
\end{eqnarray}
It is clear from this procedure that the CJT formalism eliminates the unphysical contribution to the entropy density from thermal mass corrections that we have obtained with the Coleman-Weinberg approximation to the effective potential, noting that similar gap equations can be obtained for the $\chi_{R,I}$ scalars with the identification $\lambda \leftrightarrow g^2$ (up to numerical factors) and taking into account the mixed double-bubble diagrams mentioned above, such that:
\begin{eqnarray} \label{entropy_heavy_physical}
s_X=\sum_i {\partial V\over \partial \Omega_i}{ \partial \Omega_i\over \partial T}=0~.
\end{eqnarray}
Note that a similar procedure removes the contribution to the entropy density from the thermal mass of the fermionic components, which although more involved than in the scalar case yields similar gap equations, as we describe in Appendix A. This means that the $X$ fields only contribute to the entropy density through the temperature dependence of their Bose-Einstein or Fermi-Dirac distributions, yielding the standard (positive-definite) entropy for a gas of massive bosons and fermions. In the non-relativistic regime, this contribution is Boltzmann-suppressed, and we can approximate the total entropy density by that of the light $Y$ multiplets and other potential relativistic degrees of freedom, $s =(2\pi^2/45)g_*T^3$. 

As an aside, note also that the equilibrium relation $s=(p+\rho)/T$ is recovered in the slow-roll regime, $\dot{\phi}^2/2\ll V$, with the total entropy density and pressure given in Eq ~(\ref{total_rho_p}). This is consistent since a near-equilibrium configuration can only be maintained for a slowly varying background field value.

It is interesting to observe that, even though the Coleman-Weinberg approximation considered earlier yields unphysical results for the entropy density, it is still a  very good approximation to the effective potential itself. Consider the generic scalar field example above in the non-relativistic regime, where the Boltzmann-suppressed terms may be neglected, and for a vanishing expectation value, $\langle \varphi\rangle=0$, in a closer analogy with our hybrid inflation construction. The gap equation then reads;
\begin{eqnarray} \label{gap_equation_scalar_2}
\Omega^2\left[1-{\lambda\over32\pi^2}\left(\log {\Omega^2\over\mu^2}-1\right)\right]=m^2+\alpha^2T^2~,
\end{eqnarray}
so that the corrections are suppressed for $\lambda\ll 1$ if no large logarithms are present, as should be the case for a consistent perturbative computation of the $\Phi$ function. Moreover, the effective potential becomes:
\begin{eqnarray} \label{effective_potential_scalar}
V={\lambda\over 4!}\varphi^4+{\Omega^4\over64\pi^2}\left[\log {\Omega^2\over\mu^2}-{3\over2}-{\lambda\over 32\pi^2}\left(\log {\Omega^2\over\mu^2}-1\right)^2\right]~,
\end{eqnarray}
so that the corrections to the Coleman-Weinberg term are also suppressed for $\lambda\ll1$ and not too large logarithms. What this analysis shows is that corrections such as the double-bubble diagram are indeed higher-order from the effective potential and effective mass points of view but not in computing the entropy density, being consistently taken into account within the CJT formalism described above.

Another important remark is the fact that, in computing the thermal corrections to the waterfall sector masses, we have included only the leading quantum corrections from loops of relativistic fields in the $Y$ sector, where no double-counting issues arise as we explicitly show in Appendix B. There is, however, a double-counting problem in double-bubble diagrams involving the same $\chi$ scalar fields, and the CJT formalism accounts for both this and the related unphysical contributions to the entropy density.

Given this discussion and the similar analysis for the fermionic components outlined in the appendix, it is clear that in analyzing the dynamics of SUSY warm hybrid inflation, it is a good approximation to consider the Coleman-Weinberg form for the effective potential in Eq.~(\ref{potential_xi}) including the thermal mass corrections in Eq.~(\ref{waterfall_masses_temperature}), provided that $g\ll 1$ even though we may allow for $g^2N_X\lesssim 1$, and keeping only the standard contribution from relativistic degrees of freedom to the entropy density of the system. This allows one to separate the inflaton and radiation equations of motion as in Eq.~(\ref{inflaton_equation}) and (\ref{radiation_equation}), while including the finite temperature corrections to the effective potential and dissipation coefficient.


\section{Numerical simulations}

We now have all the necessary ingredients to fully analyze the dynamics of hybrid inflation in the warm regime, taking into account (i) the full form of the dissipation coefficient including the contributions of both low-momentum and on-shell scalar $\chi_{R,I}$ modes in Eq.~(\ref{dissipation_coefficient}); and (ii) the full form of the finite temperature effective potential in the non-relativistic regime in Eq.~(\ref{potential_xi}), including the thermal corrections to the scalar and fermionic masses and SUSY splittings in the waterfall sector in Eq.~(\ref{waterfall_masses}). As concluded earlier, in the CJT improved computation of the effective potential, the latter is independent of the physical field masses, such that the waterfall sector only gives exponentially suppressed contributions to the entropy density of the system, which corresponds to a very good approximation to that of the light degrees of freedom.

It is thus possible in the general case to separate the inflaton and radiation fluid equations as in Eqs.~(\ref{inflaton_equation}) and (\ref{radiation_equation}) despite the implicit temperature dependence of the Coleman-Weinberg effective potential through the field thermal masses. It is convenient to express these equations in terms of the number of e-folds of inflation, with $dN_e=Hdt$, giving:
\begin{eqnarray} \label{inflaton_equation_efolds}
\phi''+{1\over2}\left[{V_\phi \phi'+\rho_R'\over V+\rho_R}+{\phi'\phi''\over 3M_P^2-\phi'^2/2}\right]\phi'+3(1+Q)\phi'+{V_\phi\over H^2}=0~,
\end{eqnarray}
and
\begin{eqnarray} \label{radiation_equation_efolds}
\rho_r'+4\rho_R=3Q H^2\phi'^2~,
\end{eqnarray}
where, as defined above, $Q=\Upsilon/3H$ and the Friedmann equation yields:
\begin{eqnarray} \label{Friedmann_equation_efolds}
H^2={V+\rho_R\over 3M_P^2-\phi'^2/2}~.
\end{eqnarray}
It is also straightforward to express these equations in terms of the temperature of the radiation bath for a near-equilibrium state via $\rho_R=(\pi^2/30)g_*T^4$. 

We have then performed numerical simulations of the evolution of the inflaton-radiation system from initial conditions at large field values, where the results obtained in section 3 hold. For a direct comparison with the results obtained in this regime, in Figure 3 we show the evolution of the different dynamical quantities for the same parameter choices in Figure 1, which yield $n_s=0.962$ and $r=0.019$, taking the simplest case where no other relativistic components exist besides the $Y$ sector fields, with $g_*=(15/4)N_Y$ and $\alpha^2=h^2N_Y/2$ as discussed earlier.

\begin{figure}[htbp]
\centering\includegraphics[scale=0.5]{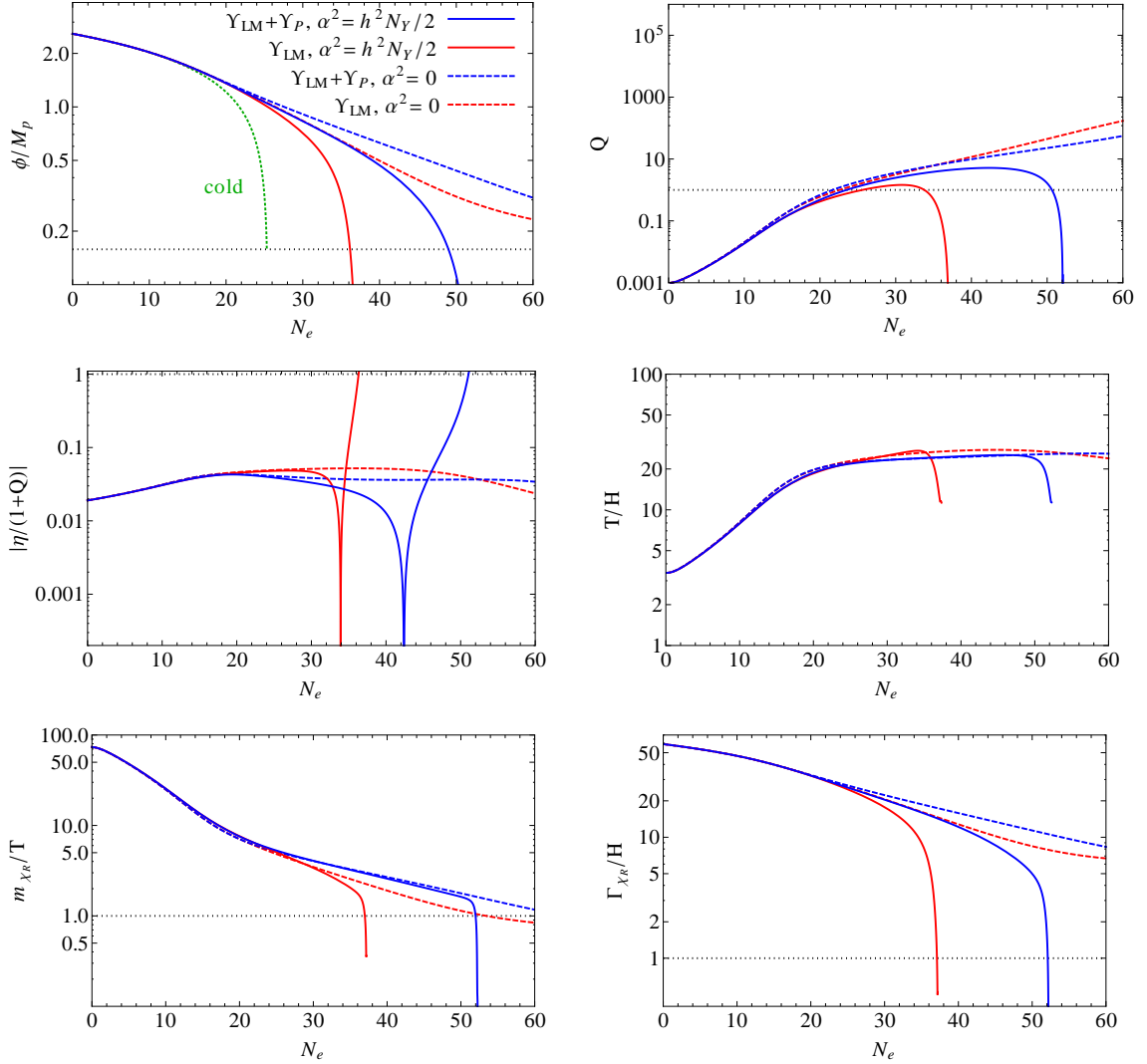}
{\vskip-0.3cm}
\caption{Numerical evolution of the different dynamical quantities using the full form of the field and radiation equations, considering only the low-momentum contribution to the dissipation coefficient (red curves) and the full dissipation coefficient including on-shell modes (blue curves), both excluding (dashed curves) and including (solid curves) thermal corrections to the waterfall sector masses. In the top left plot we also show the evolution in the supercooled regime (dotted green curve) and the chosen value of $M$ (dashed black line). These results are obtained for $g=10^{-3}$, $h=0.34$, $N_X=5\times10^6$, $N_Y=50$ and $M=0.16 M_P$, choosing $Q_*\simeq 10^{-3}$ ($\phi_{*}=2.58M_P$). For this choice of parameters, the spectral index is $n_{s}\simeq0.962$ and the tensor-to-scalar ratio is $r=0.019$, with the amplitude of the power spectrum normalized to the observational value $\Delta_\mathcal{R}^{2}\approx2.2\times10^{-9}$.}
\end{figure}

In this figure, for a better understanding of the different effects, we show the evolution of the inflaton field, temperature and related derived quantities including only the low-momentum contribution (red curves) and full dissipation coefficient (blue curves), both with (solid curves) and without (dashed curves) thermal corrections. For reference, we also include the field evolution in the corresponding supercooled regime with no dissipation and $T\ll H$, which yields approximately 25 e-folds of inflation until the waterfall transition in this case. 

Let us begin by analyzing the results when only the low-momentum dissipation coefficient is included, corresponding to the first term in Eq.~(\ref{dissipation_coefficient}). We see that, in the absence of thermal corrections, one obtains over 60 e-folds of inflation, which is due to the SUSY mass splittings the scalar $\chi_{R,I}$ components as detailed below. When thermal corrections are included, we observe that the hybrid transition, i.e. $m_{\chi_R}\rightarrow 0$, occurs for field values below the the zero-temperature critical value $\phi_c=M$, since thermal corrections increase the waterfall field masses, which should therefore prolong inflation. However, the total number of e-folds of inflation is in this case significantly smaller, with the hybrid transition occurring around 38 e-folds. This is associated with the decrease of the dissipation coefficient as a result of the larger effective masses of both the real and imaginary $X$ scalars, so that the inflaton's motion is less damped. This is easily seen by considering the next-to-leading corrections to the low-momentum dissipation coefficient in the large field limit $\phi\gg M, \alpha T/\sqrt{2}g$:
\begin{eqnarray} \label{low_momentum_dissipation_expansion}
\Upsilon_{LM}= C_\phi{T^3\over \phi^2}\left[1+10\left({M\over \phi}\right)^4-4\alpha^2\left({T\over m_X}\right)^2+\ldots\right]~,
\end{eqnarray}
where $m_X=\sqrt{2}g\phi$ is the leading scalar mass at large field values. This explicitly shows that SUSY mass splittings enhance the dissipation coefficient, while thermal corrections decrease it. Since the latter is $\mathcal{O}(T/\phi)^2$, while the former is only $\mathcal{O}(M/\phi)^4$ due to a partial SUSY cancellation, the effect of thermal corrections becomes more pronounced as one approaches the hybrid transition and the total number of e-folds is therefore reduced.

With the full form of the dissipation coefficient the results are similar, with SUSY mass splittings also enhancing (reducing) the on-shell dissipation coefficient corresponding to the real (imaginary) component of the complex $\chi$ fields, and thermal corrections equally decreasing the effect of both components. For on-shell dissipation, this effect is more pronounced since the dissipation coefficient depends exponentially on the field masses, but the added dissipation makes the field and temperature evolve more slowly, with a longer part of the evolution occurring in the strong dissipative regime, $Q\gg 1$. 

Overall, the results for the full dissipation coefficient including thermal corrections yield a total of about 52 e-folds of inflation in this example, which is therefore in agreement with hybrid inflation generating the observed primordial spectrum of density perturbations for the presently observable CMB scales. We see in this example that the slow-roll conditions are satisfied up to the hybrid transition, the same holding for the consistency conditions $T>H$, $\Gamma_{\chi_{R,I}}>H$ and $m_{\chi_{I,R}}>T$. The latter condition, in particular, is only violated very close to the transition, therefore validating our approximation of neglecting Boltzmann-suppressed contributions to the effective potential and entropy density, as well as inflaton thermal mass corrections. Note that $\Gamma_{\chi_I}>\Gamma_{\chi_R}$, so that in Figure 3 we show only the most restrictive condition. 

To better understand the evolution of the system, in Figure 4 we plot the temperature-dependent critical field value, $\phi_c(T)=\sqrt{M^2-\alpha^2T^2/2g^2}$, i.e.~the field value below which the waterfall fields become tachyonic, using the full form of the dissipation coefficient and including thermal mass corrections. 

\begin{figure}[htbp]
\centering\includegraphics[scale=0.5]{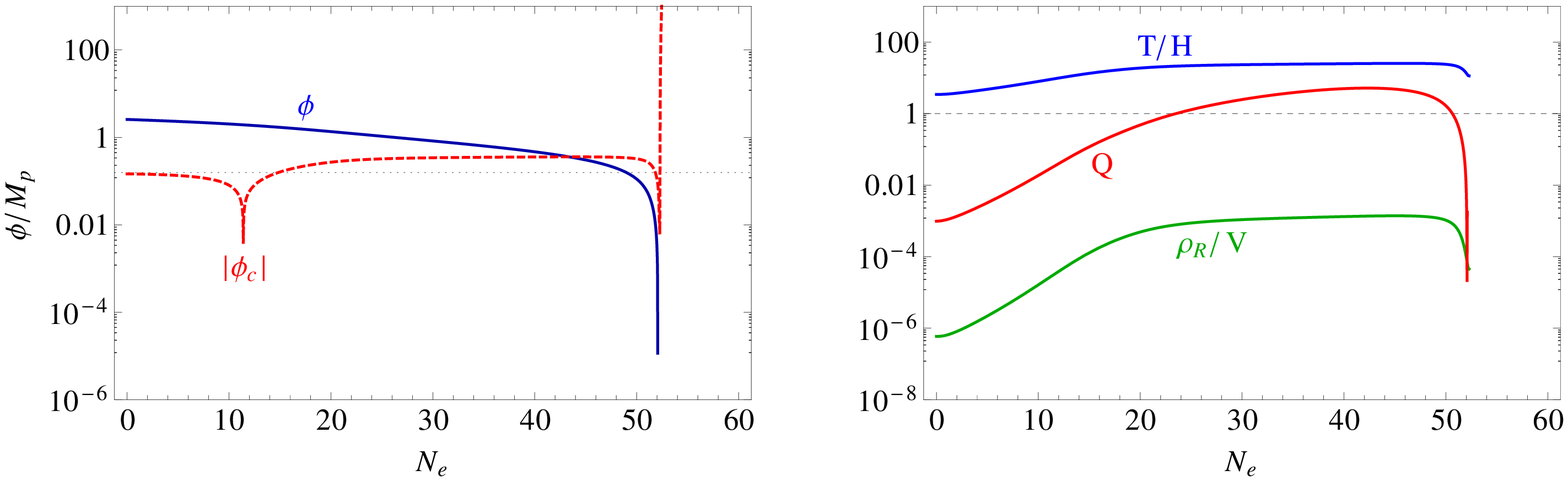}{\hskip0.25cm}
\caption{Evolution of the inflaton field compared to the finite temperature critical field value $\phi_c=\sqrt{M^2-\alpha^2T^2/2g^2}$, which reduces to $\phi_c=M$ at zero temperature (dotted black line), and is shown in absolute value (left); and evolution of the radiation abundance and related dynamical quantities (right). These results correspond to the same parameter choices as in Figure 3, using the full form of the dissipation coefficient and including thermal corrections to the field masses in the waterfall sector.}
\end{figure}

As we had seen earlier, the quantity $T/H$ increases during inflation, which in fact corresponds to an increase in the temperature of the radiation bath since the Hubble parameter does not vary significantly. This decreases the critical field value, which in this example vanishes after around 10 e-folds of accelerated expansion as shown in Figure 4. Afterwards $\phi_c$ becomes pure imaginary, signaling that the inflaton field could become arbitrarily small without triggering an instability in the waterfall sector. However, this eventually shuts down the dissipative processes, since the dissipation coefficient decreases with the value of $\phi$ according to Eq.~(\ref{dissipation_coefficient}). This explains the decrease in $Q$ observed in Figures 3 and 4, which in turn leads to a decrease in the temperature of the radiation. This makes the critical field value real once more, which ends up triggering the waterfall phase transition. This does not occur in the absence of thermal mass corrections, since in this case the waterfall field mass becomes arbitrarily small for finite inflaton field values, $\phi>M$, so that the dissipative ratio $Q$ strictly increases as shown in Figure 3.

The shutdown of dissipation due to thermal mass corrections has another important consequence, given that the subsequent cooling of the system decreases the radiation abundance towards the end of inflation, inverting the tendency observed in the absence of thermal mass corrections for radiation to smoothly take over the vacuum energy discussed in section 3. This is also illustrated in the example in Figure 4, and implies that a standard period of (p)reheating must follow the waterfall transition, with the decay of potentially both the inflaton and the waterfall fields, as well as of their superpartners, transferring the vacuum energy into relativistic degrees of freedom. Nevertheless, our simulations show that reheating will proceed from an already warm universe, since $T>H$ up to the phase transition, with radiation being subdominant but not exponentially suppressed as in supercooled scenarios. The details of the reheating period are, however, model-dependent, so that this lies outside the main focus of this work.

In Figure 5 we also show the effect of varying the effective coupling $h^2N_Y$ determining the thermal masses on the dynamical evolution. This effective coupling also modifies the strength of the low-momentum and on-shell dissipation coefficients and, in order to better identify the effects of thermal masses, we modify the basic parameters of the model so as to start from the same initial conditions, keeping $g^2N_X$, $C_\phi$, $g_*$ and $V_0$ constant. 

\begin{figure}[htbp]
\centering\includegraphics[scale=0.52]{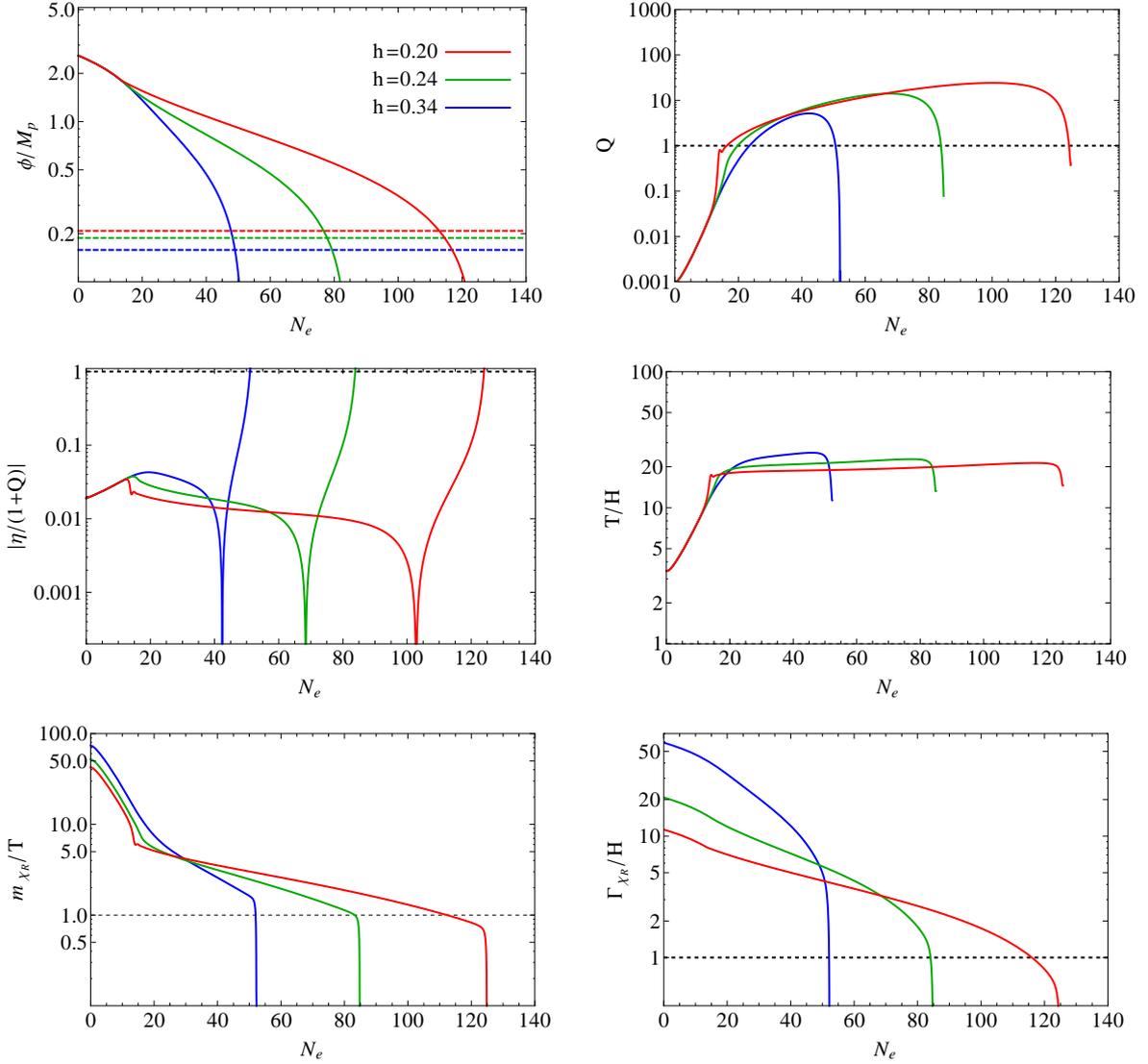}
{\vskip-0.3cm}
\caption{Numerical evolution of the different dynamical quantities using the full dissipation coefficient and including thermal mass corrections for different values of the coupling $h$, keeping $N_Y$, $g^2N_X$, $V_0$ and $C_\phi$ constant, as well as the initial inflaton value, so as to fix the same initial conditions as in Figures 2 and 3. Note that this corresponds to changing the value of the effective coupling $h^2N_Y$ and hence the effects of thermal masses and the magnitude of on-shell dissipation. The dashed horizontal lines in the top left plot give the $T=0$ critical value $\phi_c=M$ in each case.}
\end{figure}

As one can easily see in this figure, lower values of $h^2N_Y$ yield a larger number of e-folds, since as described above the effects of dissipation are less suppressed by thermal masses and, moreover, the contribution of on-shell modes is resonantly enhanced for smaller decay widths. Note that by fixing $C_\phi$ we keep the magnitude of the low-momentum contribution to the dissipation coefficient constant while varying $h^2N_Y$. Of course there is a limit on the number of e-folds one can get for lower values of $h^2N_Y$, since the rate at which the waterfall fields decay into light degrees of freedom must exceed the Hubble rate, $\Gamma_{\chi_{R,I}}>H$, or otherwise there is no light particle production and the system cannot remain near an equilibrium configuration during inflation. The examples shown in Figure 5 nevertheless show that 50-60 e-folds of inflation can be obtained in a perturbative parametric regime where this and the other consistency conditions are satisfied. One can also observe in this figure a more pronounced transition between the low-momentum and pole-dominated dissipation regimes for smaller values of $h^2N_Y$, which is associated with the fact that $\Upsilon_{LM}/\Upsilon_{P}\propto h^4N_Y$ and with the exponential dependence on the field masses for on-shell production.

In summary, our analysis shows that finite temperature and dissipative effects may significantly change the dynamics of SUSY hybrid inflation, in particular close to the waterfall transition. Dissipation helps overdamping the inflaton field's motion, therefore prolonging inflation, at the same time creating a thermal bath of relativistic particles that backreacts onto the properties of the waterfall fields and their superpartners. Thermal masses delay the hybrid transition to field values lower than the zero-temperature critical value, $\phi_c<M$, but suppress the dissipative coefficient both in the low-momentum and on-shell regimes. The dynamics of SUSY warm hybrid inflation is determined by six basic parameters, in particular the couplings $g$ and $h$, the field multiplicities $N_{X,Y}$, the fundamental mass scale $M$ and the field (or equivalently temperature) value at horizon-crossing for the relevant CMB scales. This makes a comprehensive parametric study of warm hybrid 
inflation rather complex, being outside the scope of this work, but the main dynamical features are nevertheless illustrated by the examples given above. These examples show, furthermore, that dissipative effects remove the tension between observational data and the predictions of the standard supercooled SUSY hybrid inflation scenario discussed in section 3. 


\section{Conclusion}

In this work we have analyzed the dynamics and observational predictions for supersymmetric hybrid inflation driven by radiative corrections including dissipative and associated thermal effects. Dissipation is a natural feature in these models, since the waterfall fields modify not only the effective potential but also the full effective action of the inflaton field. This leads to not only the well known logarithmic slope but also an additional damping term in the inflaton's equation of motion. This changes both the background dynamics and the evolution of field perturbations as a result of the fluctuation-dissipation relation, and we have shown that this allows for 50-60 e-folds of inflation with a primordial spectrum consistent with the results obtained by the Planck mission. This overcomes the tension in the standard supercooled scenario, where $n_s>0.98$ for $N_e>50$, which is excluded by Planck at the 95\% CL \cite{Ade:2013uln}.

Our analysis includes several novel effects that have not been considered in previous studies of warm inflation models. We have, in particular, considered the backreaction effects of the quasi-thermal bath of radiation produced by dissipation on the waterfall field masses and decay widths, showing that thermal masses modify both the effective potential and dissipation coefficient. This is particularly significant in hybrid models, since thermal corrections increase the mass of the waterfall fields, which may therefore remain in a metastable minimum with non-vanishing vacuum energy for a longer period. These same thermal mass corrections actually suppress the dissipation coefficient induced by the waterfall fields, but the balance between these two opposite effects produces nevertheless a sufficiently long period of accelerated expansion.

The inclusion of thermal mass corrections also leads to the shutdown of dissipative processes close to the end of inflation, so that despite the universe remaining in a warm phase up to the waterfall transition, radiation does not become the dominant component. This means that the transition to a radiation era will proceed through the standard perturbative decay of the inflaton and/or the waterfall fields, possibly preceded by an epoch of resonant particle production. Since, however, the radiation bath has not been exponentially redshifted during inflation, finite temperature effects will be relevant both during and after the phase transition and must be taken into account along the lines proposed in \cite{Drewes:2013iaa}. We expect this to be a particular feature of hybrid inflation models, where thermal mass corrections become relevant before the end of the inflationary phase. This is not the case, for example, of chaotic potentials \cite{Bartrum:2013fia}, for which the slow-roll conditions are generically violated at large field values and in the strong dissipation regime, where thermal mass corrections play a negligible role and radiation smoothly takes over as the dominant component.

In this work, we have also, for the first time, included the dissipative effects associated with on-shell production of waterfall field particles, which subsequently decay into light degrees of freedom in the quasi-thermal bath. Despite being Boltzmann-suppressed at the low temperatures required to avoid the generation of large inflaton masses, $\Delta m_\phi^2 \sim g^2N_XT^2\gtrsim H^2$, this contribution can in fact overcome the effect of low-momentum modes generically considered in the literature, as suggested in \cite{BasteroGil:2012cm}, and lead to a larger number of e-folds of inflation. Another effect that has been typically neglected in previous analyses is the SUSY mass splitting for chiral multiplets directly coupled to the inflaton, which as we have shown parametrically enhances the dissipation coefficient, in particular close to the hybrid transition.

An important feature in our analysis was the use of a non-perturbatively improved version of the effective potential/free energy of the system within the CJT formalism \cite{Cornwall:1974vz}, discussed in section 4. From the point of view of the effective potential itself, the corrections considered in this approach are of a higher order in the couplings and field multiplicities, having no practical effect on the dynamical evolution. However, they constitute leading corrections to the entropy density of the waterfall chiral multiplets, in particular making the effective potential independent of their physical masses, which correspond to solutions of the associated gap equations. This eliminates unphysical negative contributions to the entropy density that arise in the leading Coleman-Weinberg approximation to the effective potential. Thus, only relativistic degrees of freedom effectively contribute to the entropy of the system, which allows for the standard separation between the inflaton and radiation fluids generically considered in warm inflation models. To our knowledge, most of the discussions of the effective potential at finite temperature in the literature focus on the high temperature regime, so it is interesting to see that the CJT formalism may also be used to successfully describe the low temperature case, where as we discussed one would obtain unphysical results with the unimproved form of the thermally corrected Coleman-Weinberg potential.

 The CJT formalism is broadly used in the study of near-equilibrium systems at finite temperature, such as the quark-gluon plasma (see e.g. \cite{Blaizot:2000fc} and references therein), and we expect its range of applicability in cosmological systems to extend beyond the realm of warm inflationary dynamics. The simplest example is that of a massive scalar field coupled to a thermal radiation bath, where the (unimproved) effective potential is, to leading order, $V(\varphi,T)= m^2(T)\varphi^2/2$. Since thermal corrections generically increase the field's effective mass, $s_\varphi=-\partial V/\partial T=-(\varphi^2/2)\partial m^2/\partial T\leq 0$, i.e.~yielding a negative contribution to the entropy density away from the minimum at the origin. The improved CJT effective potential is, however, independent of the physical field mass and defines the corresponding gap equation, thus eliminating this unphysical contribution. This is of course relevant for any cosmological scenarios with scalar fields at finite temperature, such as the standard reheating phase \cite{Drewes:2013iaa, Mukaida:2012qn, Mukaida:2012bz} or cosmological phase transitions.
 
Although our analysis already takes into account several different effects that were absent in similar previous studies, it also suggests several possible extensions that can be considered in the context of hybrid inflation and other related models. The presence of dissipative effects naturally brings any system out of (local) thermal equilibrium, and near-thermal configurations can be maintained only when microphysical equilibration processes are sufficient fast, in particular compared to the Hubble expansion rate in the cosmological context. It would be interesting, however, to investigate how non-equilibrium corrections \cite{Berges:2004yj} may modify the effective action and the associated entropy density (or a generalized non-equilibrium quantity, since entropy is strictly defined in an equilibrium state). It would also be interesting to explore whether thermal mass corrections play an important role in other warm inflation models, in particular small field models such as hill-top inflation. 

An important complement to our analysis would be to consider scenarios of warm hybrid inflation with large radiative corrections, $\gamma\gtrsim 1$, where as discussed above inflaton particle production may become significant and yield a nearly-thermal state of inflaton particles, changing the scalar spectral index to the form given in Eq.~(\ref{n_s_thermal}) and discussed in \cite{Bartrum:2013fia}, so potentially compatible with observations. SUSY hybrid inflation is a natural scenario in several extensions of the Standard Model, in particular for brane-antibrane inflation \cite{Burgess:2001fx} in string/M-theory, where furthermore a large number of waterfall fields results from the D-brane multiplicity. The higher-dimensional nature of this setup changes the form of the effective potential, which is also modified by other string/compactification effects. Dissipative effects in these systems were considered in \cite{BasteroGil:2011mr}, using the large field form of the dissipation 
coefficient, and we expect similar modifications to those analyzed in this work to arise from finite temperature corrections and from the full form of the dissipation coefficient, an extension that we plan to explore in detail in the future.

This work shows explicitly that SUSY hybrid inflation is a viable model of the inflationary universe, with no need for considering scenarios with less than 50 e-folds of accelerated expansion as suggested by the Planck collaboration \cite{Ade:2013uln}, since dissipative interactions mediated by the waterfall fields can prolong inflation by the required amount. Hybrid inflation models may also provide an important link with low-energy phenomenology. Our setup, in particular, allows for a natural embedding of the MSSM fields, e.g.~in the light $Y$ sector, and the structure of the superpotential considered is ubiquitous in extensions of the Standard Model such as supersymmetric Grand Unified Theories or their analogue D-brane descriptions \cite{BasteroGil:2011mr}. We thus hope that our results motivate further exploring the embedding of the inflationary sector within a full quantum theory of fundamental particle interactions.

\acknowledgments
We would like to thank Rudnei Ramos and Ian Moss for useful discussions on this topic. AB and TPM are supported by STFC. MBG is partially supported by MICINN (FIS2010-17395) and ``Junta de Andaluc\'ia'' (FQM101). JGR is supported by FCT (SFRH/BPD/85969/2012) and partially by the grant PTDC/FIS/116625/2010 and the Marie Curie action NRHEP-295189-FP7-PEOPLE-2011-IRSES. TPM would like to acknowledge the support of SUPA and the hospitality of the University of Aveiro during the completion of this work. JGR would like to acknowledge the support and hospitality of the Higgs Centre for Theoretical Physics at the University of Edinburgh during the completion of this work.


\appendix

\section{Fermionic effective potential in the CJT formalism}

The fermionic effective potential can be computed using the CJT formalism in a similar way to the scalar component, although with some added complexity due to the momentum dependence of the fermionic self-energy as we describe below. Let us consider for simplicity a generic fermion field $\psi$ with Yukawa coupling $\lambda_f \varphi\bar\psi\psi$ to a light scalar field $\varphi$ in a thermal bath, which can be easily applied to the fermionic $\psi_\chi$ components in our setup.

Denoting by $\Sigma$ the fermionic self-energy and by $m_f$ its tree-level mass, the bare and dressed fermion propagators are given, respectively, by:

\begin{eqnarray} \label{fermion_propagators}
S_{\psi 0} = \frac{i}{p\!\!\!/-m_f}~,\qquad
S_\psi = \frac{i}{p\!\!\!/-m_f + \Sigma }~.
\end{eqnarray}
We can then write the fermionic effective potential as:
\begin{eqnarray} \label{fermion_potential}
V^{(F)}&=& i \int_p {\rm Tr}\ln S_\psi^{-1} - \int_p {\rm Tr}\Sigma S_\psi+
\Phi^{(F)}(G_\varphi,S_\psi) \,,
\end{eqnarray}
with the scalar propagator $G_\varphi$ defined as in Eq.~(\ref{scalar_propagator}). The first term gives the standard Coleman-Weinberg term and associated finite temperature correction and $\Phi^{(F)}(G_\varphi,S_\psi)$ corresponds to leading order in this case to the the ``fermionic sunset" diagram, i.e. the two-loop vacuum diagram with two fermion and one scalar propagator.

As in the scalar case, upon enforcing the stationarity of the potential with respect to the physical fermion propagator, $\delta V^{(F)}/\delta S_\psi=0$, one obtains the associated gap equation:
\begin{eqnarray} \label{fermion_gap_eq}
\frac{\delta \Phi^{(F)}} {\delta S_\psi} = \Sigma =i\int_p S_\psi G_\varphi~.
\end{eqnarray}
Solving the gap equation and getting the effective potential is not as straightforward as in the pure scalar case because the fermion self-energy $\Sigma$ depends on the external momentum, with 
\begin{eqnarray} \label{fermion_self_energy}
\Sigma(p)= \Sigma_s(p) - \gamma^0 \Sigma_0(p) + \mathbb{\gamma}\cdot \mathbf p
~\Sigma_v(p) \,.
\end{eqnarray}
which would give rise to a momentum dependent effective fermion mass, defined as:  
\begin{eqnarray} \label{fermion_mass_momentum}
M_\psi(p) = m_f - \Sigma_s(p)~.
\end{eqnarray}
Following \cite{Shu:2007vr}, we may neglect the $\Sigma_v$ term, which is found to be negligible in the Hartree-Fock approximation in studies of nuclear matter, and define the effective fermion mass as the pole of the full propagator when $\mathbf{p} \rightarrow 0$:
\begin{eqnarray} \label{fermion_mass_pole}
(p_0 -\Sigma_0(p_0))^2 = M_\psi^2 \,, 
\end{eqnarray}
so that $\Sigma_s$ and $\Sigma_0$ are evaluated taking $\mathbf{p}=0$ and setting $|p_0 -\Sigma_0|= M_\psi$. With this procedure, the fermionic effective potential can be written as:
\begin{eqnarray} \label{fermion_potential_2}
V^{(F)}&=& -\int_E\mathrm{Tr} \ln S_\psi^{-1}-{i\over2}\int_E \mathrm{Tr}\Sigma S_\psi+{i\over2}\int_E \mathrm{Tr} \Sigma^\varphi S_\psi~.
\end{eqnarray}
The first term can then be written in the form:
\begin{eqnarray} \label{fermion_potential_CW}
-\int_E\mathrm{Tr} \ln S_\psi^{-1} &=&  -\frac{1}{32 \pi^2} \Omega_\psi^4 \left(\ln\frac{\Omega_\psi^2}{\mu^2} - \frac{3}{2} \right) \nonumber\\
& & - \frac{T^4}{2 \pi^2}\int_0^\infty dx x^2 \left[ \ln ( 1 + e^{-\beta (E_F + \Sigma_0)})+ \ln ( 1 + e^{-\beta (E_F - \Sigma_0)})\right]~, 
\end{eqnarray}
where the first and second terms correspond to the vacuum and finite-temperature contributions, respectively. In the expressions above, $E_F=\sqrt{\mathbf{p}^2+M_\psi^2}$ and $\Omega_\psi^2=M_\psi^2+2(M_\psi+\Sigma_0)\Sigma_0$. The effective mass $M_\psi$ and the effective chemical potential $\Sigma_0$ then obey the following gap equations:
\begin{eqnarray} \label{fermion_gap_equations}
M_\psi &=& m_f + \lambda_f\frac{M_\psi}{16 \pi^2} \left( \ln \frac{M_\psi^2}{\mu^2} -1\right)~, \nonumber \\
& & + \frac{\lambda_f T}{2 \pi^2}\int dx \frac{x}{2 \beta E_F}\left[n_B(x)  + \frac{x}{2 \beta (E_F - M_\psi)} \left(\tilde n_+(E_F) + \tilde n_-(E_F)\right) \right] \nonumber\\
\Sigma_0 &=& \frac{\lambda_f T^2/6}{p_0} - \frac{\lambda_f T}{2 \pi^2}\int dx \frac{M_\psi + E_F}{4 M_\psi}\left(\tilde{n}_+(E_F) - \tilde n_-(E_F) \right)~,
\end{eqnarray}
where  $\tilde{n}_\pm(E_F)=\left[e^{\beta(E_F\mp \Sigma_0)}+1\right]^{-1}$ is the Fermi-Dirac distribution with an effective chemical potential $\pm \Sigma_0$.

The remaining terms in the fermion effective potential give:
\begin{eqnarray} \label{fermion_potential_3}
-{1\over2}\int_p \mathrm{Tr} \Sigma S_\psi&=&-4M_\psi (\Sigma_s I_F^+ + 2\Sigma_0 I_F^-)~,\nonumber\\
-{1\over2}\int_p \mathrm{Tr} \Sigma^\varphi S_\psi &=& 4\frac{\lambda_f T^2/6}{p_0} M_\psi I_F^- ~,
\end{eqnarray}
where we have defined the integrals:
\begin{eqnarray} \label{fermion_integrals}
I_F^{+}&=& \frac{T^2}{2 \pi^2} \int dx x^2 \frac{\tilde n_+ + \tilde n_-}{2 \beta E_F} ~,\\
I_F^{-}&=& \frac{T^2}{2 \pi^2} \int dx x^2 \frac{\tilde n_+ - \tilde n_-}{2 \beta M_\psi} ~.
\end{eqnarray}

We note that, as in the scalar case, the Coleman-Weinberg term is the leading term in the effective potential for non-relativistic fermions, $M_\psi\gg T$, and that since $p_0\simeq M_\psi$ in this limit, we have $\Omega_\psi^2= M_\psi^2+\lambda_f T^2/3$. For the $\psi_\chi$ fermion in our hybrid setup, this corresponds to the Coleman-Weinberg contribution used in Eq.~(\ref{potential_xi}) with thermal mass corrections identical to its scalar superpartners $\chi_{R,I}$ in the waterfall sector. Nevertheless, the additional terms introduced in the CJT formalism are not sub-leading for computing the entropy density and remove the unphysical contribution from the temperature dependence of the effective fermion mass.


\section{Resummation and the double-counting problem}

Consider a scalar field of tree-level mass $m$ and quartic self-coupling $\lambda\varphi^4/4!$ at finite temperature, such that $T\gg m$. The effective mass of the scalar, assuming a vanishing expectation value for simplicity, is then $\bar{m}^2=m^2+\lambda T^2/24$. Resummation of the corresponding daisy and superdaisy diagrams then results in the replacement of the tree-level scalar mass by this effective mass in the scalar propagators, modifying the effective potential. This modifies in particular the vacuum (Coleman-Weinberg) part of the effective potential, which is given by:
\begin{eqnarray} \label{scalar_CW}
V^{(vac)}={1\over 2}\int {d^4p\over (2\pi)^4}\log \left(p^2+m^2+{\lambda\over 24}T^2\right)~.
\end{eqnarray}
Expanding this in powers of the self-coupling $\lambda$ we obtain to leading order:
\begin{eqnarray} \label{scalar_CW_expansion}
V^{(vac)}=V^{(vac)}(T=0)+{\lambda\over 48}T^2\int {d^4p\over (2\pi)^4}{1\over p^2+m^2}+\ldots
\end{eqnarray}
The double-bubble diagram with two $\varphi$-loops is equivalent to a single bubble diagram with a thermal mass insertion $\lambda T^2/24$, with a symmetry factor $1/2$ for a single bubble and an additional $1/2$ factor accounting for the identical fields in the two loops. This gives: 

\begin{eqnarray}
\fcolorbox{white}{white}{
  \begin{picture}(313,50) (40,-40)
    \SetWidth{1.0}
    \SetColor{Black}
    \Arc(10,0)(20,54,420)
    \Arc(50,0)(20,54,420)
    \Text(85,-6)[lb]{\Black{$={1\over2}$}}
     \Arc(130,0)(20,54,420)
     \Vertex(130,20){2.627}
      \Text(160,-8)[lb]{\Black{$={1\over2}\times \frac{\lambda}{24}T^2\times {1\over2}\int {d^4p\over (2\pi)^4}{1\over p^2+m^2}={\lambda\over 96}T^2\int {d^4p\over (2\pi)^4}{1\over p^2+m^2}$~,}}
  \end{picture}
}
\end{eqnarray}
which clearly shows that this diagram appears twice in the expansion of the effective potential with the thermal mass correction. 

If the field $\varphi$ is coupled to another light field in the thermal bath, e.g. through an operator $\lambda_\sigma \varphi^2\sigma^2$, the effective mass is also corrected by a factor $\Delta m^2= \lambda_\sigma T^2/6$. We can proceed as above to compute the effective potential with the new effective mass and again expand the result in powers of $\lambda_\sigma$. This is entirely analogous to the result above, except for the fact that the relevant double-bubble diagram now involves different fields, which removes one of the $1/2$ symmetry factors in the expression above. Hence, the expansion of the effective potential gives the double-bubble diagram to leading order, and there is no double-counting to leading order in this case.


\bibliographystyle{JHEP}

\bibliography{warmhybrid_refs}

\end{document}